\def\tr{{\text{tr}}\,}
\def\Tr{{\text{Tr}}\,}
\def\wt{\widetilde}

\def\be{\begin{equation}}
\def\ee{\end{equation}}
\def\bea{\begin{eqnarray}}
\def\eea{\end{eqnarray}}
\def\bse{\begin{subequations}}
\def\ese{\end{subequations}}
\def\b{\bibitem}
\documentclass[prb,twocolumn,showpacs,amsmath,amssymb,eqsecnum]{revtex4} 
\usepackage{graphicx}
\usepackage{dcolumn}
\usepackage{bm}
\begin{document}
\title{Quantum Metal--Superconductor Transition:
A Local Field Theory Approach
}
\author{Lubo Zhou}
\affiliation{Institute for Physical Science and Technology\\
University of Maryland,\\ 
College Park, MD 20742}
\author{T.R. Kirkpatrick}
\affiliation{Institute for Physical Science and Technology, and Department of 
         Physics\\
         University of Maryland,\\ College Park, MD 20742}
\date{\today}

\begin{abstract}

The zero temperature, or quantum, metal-superconductor phase transition is studied 
in disordered systems in dimension greater than two. A effective 
local field theory is developed that keeps all soft modes or fluctuations explicitly. 
A simple renormalization group 
analysis is used to exactly determine the quantum critical behavior at this transition.

\end{abstract}



\maketitle

\section{Introduction}
\label{sec:I}

Recently there has been much interest in quantum phase transitions. 
Occurring at $T=0$, these transitions provide new insight into 
the possible physical phases of systems at  low temperature.\cite{Sachdev_book} The first 
quantum phase transition studied in detail was the ferromagnetic transition in itinerant 
electron system at zero temperature. Hertz argued in 1976 that the transition was mean-field-like 
in the physically interesting dimension $d=3$.\cite{Hertz} This simple mean-field description 
was later shown to be incorrect.\cite{Sachdev,us_dirty,us_clean} The reason for this breakdown is the 
existence of soft or massless modes other than the order parameter fluctuations. These modes were 
being neglected in Hertz's theory. In disordered systems 
these modes are diffusive, and they couple to the order parameter
fluctuations and modify the critical behavior.\cite{us_dirty,us_clean}
Technically, if these additional soft modes are 
integrated out, they lead to
a long-ranged interaction and a nonlocal field theory. It was
argued that once this effect is taken into account, all other fluctuation
effects are suppressed by the long-range nature of the interactions and that
the critical behavior is governed by a fixed point that is Gaussian, but
does not yield mean-field exponents.

Similar argument was used to describe the normal metal-to-superconductor quantum phase transition at $T=0$.\cite{MSC}
In this case the usual finite temperature superconducting 
phase transition is driven to zero temperature by nonmagnetic 
disorder,\cite{small} where the additional soft modes come from particle-hole excitations.
Again, it was argued that the critical behavior found at 
this quantum phase transition\cite{MSC} could be exactly determined using the 
same technique as in Refs.\ \onlinecite{us_dirty,us_clean}.


The theory developed in Refs.\ \onlinecite{us_dirty,us_clean,MSC}, however,  suffered from one
major drawback: Since the additional soft modes were integrated out in order
to obtain a description entirely in terms of the order parameter fluctuations, the effective
field theory that was derived was nonlocal\cite{local_footnote}
and not suitable for
perturbatively calculating effects that depend on all of the soft modes in the
system. The analysis in 
Refs.\ \onlinecite{us_dirty,us_clean,MSC} was therefore restricted to
power counting arguments
at tree level to show that all non-Gaussian terms are irrelevant in a 
RG sense. However, relying entirely
on tree-level power counting can be dangerous. 
In particular, logarithmic corrections can be easily missed.
Later on logarithmic corrections were indeed found in the description of the 
quantum ferromagnetic transition.\cite{us_paper_III}

It is the motivation of this paper to keep all the relevant soft modes and to construct an 
effective local field theory for the metal-superconductor transition so that the exact behavior  at 
this quantum phase transition can be determined. 
Unlike the quantum ferromagnetic transition discussed above, we will show that the previous results 
for the metal-superconductor
transition, though from nonlocal field theory treatment, are still valid. The inherent reason for that
is explained in detail.

This paper is organized as follows. In Sec.\ \ref{sec:II} we use methods
developed in Refs.\ \onlinecite{us_fermions,LZ} to derive an
effective local theory for disordered electron systems that
explicitly
 separates massive modes from soft ones, and 
keeps all of the latter. In Sec.\ \ref{sec:III} we give a renormalization group analysis
of this model. In Sec.\ \ref{sec:IV}
we discuss our results.


\section{Effective Local Field Theory}
\label{sec:II}

A local field theory will be given in this section to describe the normal metal-to-superconductor 
quantum phase transition at $T=0$. All relevant soft modes will be contained in this field theory. 
We start from a general model of interacting electrons with quenched disorder and an attractive Cooperon
interaction amplitude. We then introduce 
the superconducting order parameter and separate massive and soft modes. After integrating out 
the massive modes, we obtain a effective local field theory that describe the coupling between 
the superconducting fluctuations and the soft or massless diffusive modes.

\subsection{Composite field theory}
\label{subsec:II.A}

The general partition function of the interacting, disordered electrons can be given in 
the form of Grassmann fields $\bar\psi$ and $\psi$\cite{NegeleOrland} 
\bse
\label{eqs:2.1}
\begin{equation}
Z=\int D[\bar{\psi},\psi]\ e^{S[\bar{\psi},\psi]}\quad. 
\label{eq:2.1a}
\end{equation}
with the action $S$ being 
\bea
S &=& - \int_0^{\beta} d\tau \int d{\bf x} \sum_{\sigma}\ 
  {\bar\psi}_{\sigma}({\bf x},\tau)\,{\frac{\partial}{\partial\tau}}\,\psi_{\sigma}({\bf x},\tau)
\nonumber\\
 && - \int_0^{\beta} d\tau\ H(\tau)\ .
\label{eq:2.1b}
\eea
\ese%
We denote the spatial position by ${\bf x}$, and the imaginary time by $\tau$.
$H(\tau)$ is the Hamiltonian in imaginary time representation,
$\beta=1/T$ is the inverse temperature, $\sigma=1,2$ denotes spin labels.
As what we have done in previous papers, we integrate out the Grassmann fields and rewrite the 
theory in terms of complex-number fields $Q$ and ${\wt\Lambda}$. With the help of the following isomorphism,
\begin{eqnarray}
B_{12} &=& \frac{i}{2}\,\left( \begin{array}{cccc}
          -\psi_{1\uparrow}{\bar\psi}_{2\uparrow} &
             -\psi_{1\uparrow}{\bar\psi}_{2\downarrow} &
                 -\psi_{1\uparrow}\psi_{2\downarrow} &
                      \ \ \psi_{1\uparrow}\psi_{2\uparrow}  \\
          -\psi_{1\downarrow}{\bar\psi}_{2\uparrow} &
             -\psi_{1\downarrow}{\bar\psi}_{2\downarrow} &
                 -\psi_{1\downarrow}\psi_{2\downarrow} &
                      \ \ \psi_{1\downarrow}\psi_{2\uparrow}  \\
          \ \ {\bar\psi}_{1\downarrow}{\bar\psi}_{2\uparrow} &
             \ \ {\bar\psi}_{1\downarrow}{\bar\psi}_{2\downarrow} &
                 \ \ {\bar\psi}_{1\downarrow}\psi_{2\downarrow} &
                      -{\bar\psi}_{1\downarrow}\psi_{2\uparrow} \\
          -{\bar\psi}_{1\uparrow}{\bar\psi}_{2\uparrow} &
             -{\bar\psi}_{1\uparrow}{\bar\psi}_{2\downarrow} &
                 -{\bar\psi}_{1\uparrow}\psi_{2\downarrow} &
                      \ \ {\bar\psi}_{1\uparrow}\psi_{2\uparrow} \\
                    \end{array}\right)
\nonumber\\
&\cong& Q_{12}\quad,
\label{eq:2.3}
\end{eqnarray}
we exactly rewrite the partition 
function as\cite{us_fermions,LZ}
\begin{eqnarray}
Z &=& \int D[{\bar\psi},\psi]\ e^{{\tilde S}[{\bar\psi},\psi]}
      \int D[Q]\,\delta[Q-B]
\nonumber\\
  &=& \int D[{\bar\psi},\psi]\ e^{{\tilde S}[{\bar\psi},\psi]}
      \int D[Q]\,D[{\wt\Lambda}]\ e^{\Tr [{\wt\Lambda}(Q-B)]}
\nonumber\\
  &\equiv& \int D[Q]\,D[{\wt\Lambda}]\ e^{{\cal A}[Q,{\wt\Lambda}]}\quad.
\label{eq:2.4}
\end{eqnarray}
Here ${\wt\Lambda}$ is an auxiliary bosonic matrix field that plays the role of 
a Lagrange multiplier. The reason to do so is that the rewritten action is 
particularly suited for the separation of massive and soft modes. We then decouple 
the particle-particle spin-singlet interaction by means of a
Hubbard-Stratonovich transformation. Denoting the Hubbard-Stratonovich field by $\Psi$, 
the partition function becomes 
\bse
\label{eqs:2.5}
\be
Z = \int D[Q,{\wt\Lambda},\Psi]\ e^{\wt{\cal A}[Q,{\wt\Lambda},\Psi]}\quad,
\label{eq:2.5a}
\ee
where the action
\bea
\wt{\cal A}[Q,{\wt\Lambda},\Psi]&=&{\cal A}_{\rm dis}[Q] 
   + {\cal A}_{\rm int}^{(s)}[Q] 
+ {\cal A}_{\rm int}^{(t)}[Q] 
\nonumber\\
   &&+ \frac{1}{2}\,\Tr\ln\left(G_0^{-1} - i{\wt\Lambda}\right)
 + \Tr\left({\wt\Lambda}Q\right)
\nonumber\\
   &&- \int d{\bf x}\sum_{\alpha}\sum_n\sum_{r=1,2} {_r\Psi}_n^{\alpha}({\bf x})\,
          {_r\Psi}_{n}^{\alpha}({\bf x})
\nonumber\\
   &&+ i\sqrt{2T\vert\Gamma^{(c)}\vert}\int d{\bf x}
                              \sum_{\alpha}\sum_n\sum_{r=1,2}
          {_r\Psi}_n^{\alpha}({\bf x})
\nonumber\\
   &&\times\sum_m \tr\left[\left(\tau_r\otimes s_0\right)\,
                   Q_{m,-m+n}^{\alpha\alpha}({\bf x})\right]\ .
\nonumber\\
\label{eq:2.5b}
\eea
with $\Tr$ denoting a trace over all degrees of freedom, including the
continuous real space position, and $\tr$ a trace over all discrete
degrees of freedom that are not summed over explicitly. $\Gamma^{(c)} <0 $ is the attractive Cooperon interaction amplitude.
The first three
terms in Eq.\ (\ref{eq:2.5b}) have the following forms,
\be 
{\cal A}_{\rm dis}[Q] = \frac{1}{\pi N_{\rm F}\tau_{\rm e}}\int d{\bf x}\ 
   \tr \left(Q({\bf x})\right)^2\quad,
\label{eq:2.5c}
\ee
\bea
{\cal A}_{\rm int}^{\,(s)}&=&\frac{T\Gamma^{(s)}}{2}\int d{\bf x}\sum_{r=0,3}
   (-1)^r \sum_{n_1,n_2,m}\sum_\alpha
\nonumber\\
&&\times\left[\tr \left((\tau_r\otimes s_0)\,Q_{n_1,n_1+m}^{\alpha\alpha}
({\bf x})\right)\right]
\nonumber\\
&&\times\left[\tr \left((\tau_r\otimes s_0)\,Q_{n_2+m,n_2}^{\alpha\alpha}
({\bf x})\right)\right]\quad,
\label{eq:2.5d}
\eea
\bea
{\cal A}_{\rm int}^{\,(t)}&=&\frac{T\Gamma^{(t)}}{2}\int d{\bf x}\sum_{r=0,3}
   (-1)^r \sum_{n_1,n_2,m}\sum_\alpha\sum_{i=1}^{3}
\nonumber\\
&&\times\left[\tr \left((\tau_r\otimes s_i)\,Q_{n_1,n_1+m}^{\alpha\alpha}
({\bf x})\right)\right]
\nonumber\\
&&\times\left[\tr \left((\tau_r\otimes s_i)\,Q_{n_2+m,n_2}^{\alpha\alpha}
({\bf x})\right)\right]\quad,
\label{eq:2.5e}
\eea
Finally, 
\be 
G_0^{\,-1} = -\partial_{\tau} +\nabla^2 /2m+\mu \quad,
\label{eq:2.5f}
\ee
\ese%
is the inverse Green operator. 

Physically, the Hubbard-Stratonovich field $\Psi$ can be related to the superconducting, or Cooper pair, order parameter, $\Psi\sim \psi\psi$.

\subsection{Soft modes}
\label{subsec:II.B}

Now we are ready to separate the massive and soft modes. Our calculations 
follow the same procedure in previous papers.\cite{us_fermions} Here we just simply quote
the results. The effective action has the form
\bea
{\cal A}[q,\Psi,\Delta P,\Delta\Lambda]&=&{\cal A}_{{\rm NL}\sigma{\rm M}}[q]
   + \delta{\cal A}[\Delta P,\Delta\Lambda,q]
\nonumber\\
 &&\hskip -30pt - \int d{\bf x}\sum_{\alpha}\sum_n\sum_{r=1,2} 
     {_r\Psi}_n^{\alpha}({\bf x})\, {_r\Psi}_{n}^{\alpha}({\bf x})
\nonumber\\
 &&\hskip -30pt + i\sqrt{\pi T\vert K^{(c)} \vert}\int d{\bf x}
                              \sum_{\alpha}\sum_n\sum_{r=1,2} {_r\Psi}_n^{\alpha}({\bf x})
\nonumber\\
   &&\hskip -30pt \times\sum_m \tr\left(\tau_r\otimes s_0\right)\,
      \left[{\hat Q}^{\alpha\alpha}_{m,-m+n}({\bf x}) \right.
\nonumber\\
       &&\hskip -30pt + \left.\frac{4}{\pi N_{\rm F}}\,
      \left({\cal S}\Delta P{\cal S}^{-1}\right)
              ^{\alpha\alpha}_{m,-m+n}({\bf x})\right]\ .
\nonumber\\
\label{eq:2.6}
\eea
Here $K^{(c)} = \pi N_{\rm F}^2\Gamma^{(c)}/8$.  
The matrix $\cal S$ can be treated as ${\cal S}=1$ when we neglecting 
the coupling
between massless modes $q$ and these massive fluctuations $\Delta P$ and
$\Delta\Lambda$, which is irrelevant for the purpose of the current paper.

${\cal A}_{{\rm NL}\sigma{\rm M}}$ is the known action of the nonlinear
sigma model,\cite{NL}
\bse
\label{eqs:2.7}
\bea
{\cal A}_{{\rm NL}\sigma{\rm M}}&=&{\cal A}_{\rm int}^{(s)}[\pi\,N_{\rm F}{\hat Q}/4]
 + {\cal A}_{\rm int}^{(t)}[\pi\,N_{\rm F}{\hat Q}/4]
\nonumber\\
&&+
\frac{-1}{2G}\int d{\bf x}\
     \tr\left(\nabla\hat Q ({\bf x})\right)^2
\nonumber\\
&&+ 2H \int d{\bf x}\ \tr\left(\Omega\,{\hat Q}({\bf x})
     \right)\quad,	
\label{eq:2.7a}
\eea
with ${\cal A}_{\rm int}^{(s)}$ from Eq.\ (\ref{eq:2.5d}), 
${\cal A}_{\rm int}^{(t)}$ from Eq.\ (\ref{eq:2.5e}). 
and $\Omega$ a frequency matrix with elements
\be
\Omega_{12} = \left(\tau_0\otimes s_0\right)\,\delta_{12}\,\omega_{n_1}\quad.
\label{eq:2.7b}
\ee
Here 
$\hat Q$ is subject to the following constraints,
\begin{equation}
{\hat Q}^2({\bf x}) \equiv \openone\otimes\tau_0\quad,\quad 
       {\hat Q}^{\dagger} = {\hat Q}\quad,\quad\tr{\hat Q}({\bf x}) = 0\quad.
\label{eq:2.7c}
\end{equation}
and then can be write in a block matrix form as
\be
{\hat Q} = \left( \begin{array}{cc}
                 \sqrt{1-qq^{\dagger}} & q   \\
                    q^{\dagger}        & -\sqrt{1-q^{\dagger} q} \\
           \end{array} \right)\quad,
\label{eq:2.7d}
\ee
\ese%
where the matrix $q$ has elements $q_{nm}$ whose frequency labels are
restricted to $n\geq 0$, $m<0$. Symmetry analysis with Ward identities ensures that 
the matrix $q$ are massless, which are diffusive in disordered systems.
The coupling constants $G$ and $H$ are proportional to the inverse
conductivity, $G\propto 1/\sigma$, and the specific heat coefficient,
$H\propto\gamma\equiv\lim_{T\rightarrow 0}\,C_V/T$,
respectively.\cite{us_R,CC}

$\delta{\cal A}$ contains the corrections to the nonlinear sigma model
that were given in Ref.\ \onlinecite{us_fermions}. We list explicitly
the terms that are bilinear in the massive fluctuations $\Delta P$ and
$\Delta\Lambda$, but do not contain couplings between
the massive modes and $q$,
\bea
\delta{\cal A}^{(2)}&=&{\cal A}_{\rm dis}[\Delta P]
   + \int d{\bf x} \tr\left(\Delta\Lambda ({\bf x})
     \Delta P ({\bf x})\right) 
\nonumber\\
   &&\hskip -30pt + \frac{1}{4}\int d{\bf x} d{\bf y}
      \tr\bigl(G({\bf x}-{\bf y})\,
   \Delta\Lambda({\bf y})\,G({\bf y}-{\bf x})\,\Delta\Lambda ({\bf x})\bigr)
\nonumber\\
\label{eq:2.8}
\eea
with ${\cal A}_{\rm dis}^{(s)}$ from Eq.\ (\ref{eq:2.5c}).

Let us first integrate out $\Delta P$ and
$\Delta\Lambda$ in a Gaussian approximation. A additional quadratic contribution in terms of 
the order parameter field $\Psi$ will obtained from Eq.\ (\ref{eq:2.8}) 
and the last term in
Eq.\ (\ref{eq:2.6}). 
Combining it
with the $\Psi^2$ term in Eq.\ (\ref{eq:2.6}) yields a term 
\bse
\label{eqs:2.9}
\bea
{\cal A}_{\rm G}[\Psi] &=& -\sum_{\bf k}\sum_{\alpha}\sum_n\sum_{r=1,2}{_r\Psi}_n^{\alpha}({\bf k})
\nonumber\\
&&\times
   \left[1 + 2\Gamma_{(c)}{\tilde\chi}({\bf k},\Omega_n)\right]\,
       {_r\Psi}_{n}^{\alpha}(-{\bf k})\ ,
\label{eq:2.9a}
\eea
where
\be
{\tilde\chi}({\bf k},\Omega_n) = T\sum_{n_1,n_2}\,\Theta(n_1n_2)\,
   \delta_{n_1+n_2,n}\,{\cal D}_{n_1n_2}({\bf k})\quad,
\label{eq:2.9b}
\ee
is given in terms of
\be
{\cal D}_{nm}({\bf k}) = \varphi_{nm}({\bf k})\,
   \left[1 - \frac{1}{2\pi N_{\rm F}
   \tau_{\rm el}}\,\varphi_{nm}({\bf k})\right]^{-1}
\label{eq:2.9c}
\ee
with
\be
\varphi_{nm}({\bf k}) = \frac{1}{V}\sum_{\bf p} G_{\rm sp}({\bf p},\omega_n)\,
   G_{\rm sp}({\bf p}+{\bf k},\omega_m)\quad.
\label{eq:2.9d}
\ee
\ese%
Here $G_{\rm sp}$ is the saddle-point Green function from the inverse of 
Eq.\ (\ref{eq:2.5f}). The Theta-function in Eq.\ (\ref{eq:2.9b}),
which restricts the frequency sum to frequencies that both have the same sign.
For small frequencies 
and wavenumbers, the calculation shows that
\bse
\label{eqs:2.10}
\be
{\cal A}_{\rm G}[\Psi] = -\sum_{\bf k}\sum_{\alpha}\sum_n\sum_{r=1,2} {_r\Psi}_n^{\alpha}({\bf k})\,
   u_2\,
       {_r\Psi}_{n}^{\alpha}(-{\bf k})\quad, 
\label{eq:2.10a}
\ee
with
\be
u_2 = 1 + O({\bf k}^2,\Omega_n)\quad.
\label{eq:2.10b}
\ee
\ese%
Below we will see the wavenumber and frequency corrections indicated in Eq.\ (\ref{eq:2.10b}) are irrelevant for the critical behavior.

Now we can write an effective local action including only soft modes 
and the superconducting order-parameter fluctuations. The action has the form of
\bse
\label{eqs:2.11}
\be
{\wt{\cal A}}_{\rm eff}= {\cal A}_{\rm G}[\Psi] + {\cal A}_{{\rm NL}\sigma{\rm M}}[q]
+ {\cal A}_{\rm c}[\Psi, q]\quad.
\label{eq:2.11a}
\ee
Here the nonlinear sigma model part of the action, 
${\cal A}_{{\rm NL}\sigma{\rm M}}$, has been given in Eqs.\ (\ref{eqs:2.7}), 
and ${\cal A}_{\rm c}$ represents the coupling between $\Psi$ and $q$,
\bea
{\cal A}_{\rm c}[\Psi,q]&=&i\sqrt{\pi T\vert K^{(c)}\vert}\int \hskip -2pt d{\bf x}
                              \sum_{\alpha}\sum_n\sum_{r=1,2}
         {_r\Psi}_n^{\alpha}({\bf x})\hskip -2pt 
\nonumber\\
&&\times\sum_m\tr\left[\left(\tau_r\otimes s_0\right)\,
      {\hat Q}^{\alpha\alpha}_{m,-m+n}({\bf x}) \right] .
\nonumber\\
\label{eq:2.11b}
\eea
For the simplicity, we rewrite the coupling action as
\be
{\cal A}_{\rm c}[b,q] = i\sqrt{\pi T\vert K^{(c)}\vert}\int d{\bf x}\ \tr
   \left(b({\bf x})\,{\hat Q}({\bf x})\right)\quad.
\label{eq:2.12}
\ee
\ese%
Here we define a field
\bse
\label{eqs:2.13}
\be
b_{12}({\bf x}) = \sum_{r=1,2} (\tau_r\otimes s_0)\,{_rb}_{12}({\bf x})\quad,
\label{eq:2.13a}
\ee
with components
\bea
{_rb}_{12}({\bf k})&=&\delta_{\alpha_1\alpha_2} \sum_n
     \delta_{n,n_1+n_2} {_r\Psi}_n^{\alpha_1}({\bf k})\quad.
\label{eq:2.13b}
\eea
\ese%
Higher-order corrections can be shown be irrelevant the same way as in Ref.\ \onlinecite{us_paper_III}
Using Eq.\ (\ref{eq:2.7d}) in Eq.\ (\ref{eq:2.12}),
and integrating out the massive $P$-fluctuations, obviously
leads to a series of terms coupling $\Psi$ and $q$, $\Psi$
and $q^2$, etc. We thus obtain ${\cal A}_{\rm c}[\Psi, q]$ in form of a series
\bse
\label{eqs:2.58}
\be
{\cal A}_{\rm c}[\Psi, q] = {\cal A}_{\Psi-q} + {\cal A}_{\Psi-q^2} + \ldots
\label{eq:2.58a}
\ee
The first term in this series is obtained by just
replacing $Q$ by $q$ in Eq.\ (\ref{eq:2.12}),
\be
A_{\Psi-q} = i c_1T^{1/2}\int d{\bm x}\ \tr\left(b({\bm x})\,q({\bm x})\right) 
\label{eq:2.58b}
\ee
with $c_1=\sqrt{\pi\vert K^{(c)}\vert}$. The next term in this expansion yields
\be
A_{\Psi-q^2} = i c_{2}\sqrt{T} \int d{\bm x}\ \tr \left(b({\bm x})\,q({\bm x})\,
   q^{\dagger}({\bm x})\right) 
\label{eq:2.58c}
\ee
\ese%
with $c_2=c_1/16$. Terms of higher order in $q$
in this expansion will turn out to be irrelevant for determining the critical
behavior at the quantum phase transition.

\section{Renormalization group analysis}
\label{sec:III}

\subsection{Gaussian Propagators}
\label{subsubsec:III.A}

For the purpose of renormalization group analysis, we first need to determine the behaviors of 
Gaussian Propagators. The Gaussian or second-order action can be written as follows,
\bse
\label{eqs:2.14}
\bea
{\cal A}^{(2)}[\Psi,q]&=&-\sum_{\bf k}\sum_n\sum_{\alpha}\sum_{r=1,2}
   {_r\Psi}_n^{\alpha}({\bf k})\,u_2({\bf k})\,{_r\Psi}_{n}^{\alpha}(-{\bf k})
\nonumber\\
&&\hskip -30pt - \frac{4}{G}\sum_{\bf k}\sum_{1,2,3,4}\sum_{i,r}{^i_rq}_{12}({\bf k})\,
        {^i_r\Gamma}_{12,34}^{(2)}({\bf k})\,{^i_rq}_{34}(-{\bf k})
\nonumber\\
&&\hskip -30pt - 8 i \sqrt{\pi T\vert K^{(c)}\vert}\sum_{\bf k}\sum_{12}\sum_{i,r}{^i_rq}_{12}({\bf k})\,
    {^i_rb}_{12} (-{\bf k}) ,
\nonumber\\
\label{eq:2.14a}
\eea
where the bare two-point $q$ vertex has the form
\bea
{^0_{1,2}\Gamma}_{12,34}^{(2)}({\bf k})&=&- \delta_{13}\delta_{24}\left({\bf k}^2
   + GH\Omega_{n_1-n_2}\right) + \delta_{1+2,3+4}
\nonumber\\
&&\times\delta_{\alpha_1\alpha_2}
     \delta_{\alpha_1\alpha_3}\,4\pi TG\delta{k_c} ,
\label{eq:2.14b}\\
{^0_{0,3}\Gamma}_{12,34}^{(2)}({\bf k})&=&\delta_{13}\delta_{24}\left({\bf k}^2
   + GH\Omega_{n_1-n_2}\right) + \delta_{1-2,3-4}
\nonumber\\
&&\times\delta_{\alpha_1\alpha_2}
     \delta_{\alpha_1\alpha_3}\,4\pi TGK_s ,
\label{eq:2.14c}\\
{^{1,2,3}_{0,3}\Gamma}_{12,34}^{(2)}({\bf k})&=&\delta_{13}\delta_{24}\,\left(
   {\bf k}^2 + GH\Omega_{n_1-n_2}\right) + \delta_{1-2,3-4}
\nonumber\\
&&\times\delta_{\alpha_1\alpha_2}
     \delta_{\alpha_1\alpha_3}\,4\pi TGK_t ,
\label{eq:2.14d}
\eea
\ese%
with $K_s = -\pi\,N_{\rm F}^2\,\Gamma^{(s)}/8$ and $K_t = -\pi\,N_{\rm F}^2\,\Gamma^{(t)}/8$. Note that there is an 
additional repulsive interaction, $\delta{k_c}$, in Eq.\ (\ref{eq:2.14b}), which comes from the one-loop renormalization 
of the action.\cite{small} We choose to take this effect into account at Gaussian order. Alternately, it would arise as a higher order disorder effect.
For a complete discussion of this term we refer elsewhere.\cite{us_R}
Here we note that it is this term that drives the superconducting 
transition temperature to zero, and leads to a quantum metal-superconductor 
phase transition.

If the fermion $q$ fields are integrated out, an effective action 
including only the superconducting order parameter can be obtained. In the long wavelength and low frequency limit,
\bea
{\cal A}^{(2)}[\Psi] &=& -\sum_{\bf k}\sum_n\sum_{\alpha}\sum_{r=1,2}
   {_r\Psi}_n^{\alpha}({\bf k})
\nonumber\\
&&\left(u_2({\bf k}) 
+  
\frac{\frac{-\vert K^{(c)}\vert}{H} \ln{\frac{\Omega_0}{\vert\Omega_n\vert + 
{\bf k}^2/GH}}}
{1 + \frac{\delta{k_c}}{H} \ln{\frac{\Omega_0}{\vert\Omega_n\vert + 
{\bf k}^2/GH}}}\right)
\,{_r\Psi}_{n}^{\alpha}(-{\bf k})
\nonumber\\
&\simeq&-\sum_{\bf k}\sum_n\sum_{\alpha}\sum_{r=1,2}
   {_r\Psi}_n^{\alpha}({\bf k})
\nonumber\\
&&\left(t + \frac{\vert K^{(c)}\vert}{\delta{k_c}^2}\frac{1}{\ln{\frac{\Omega_0}{\vert\Omega_n\vert + 
{\bf k}^2/GH}}}
\right)
\,{_r\Psi}_{n}^{\alpha}(-{\bf k})\quad.
\nonumber\\
\label{eq:2.50}
\eea
Here $\Omega_0$ is a frequency cutoff on the order of the Debye frequency, and $t = u_2 - \frac{\vert K^{(c)}\vert}{\delta{k_c}}$ denotes the distance 
from the mean field or Gaussian 
critical point. ${\cal A}^{(2)}[\Psi]$ is the Gaussian order parameter field theory that was considered in Ref.\ \onlinecite{MSC}.

For the coupled field theory it is straightforward to compute the two-point correlation functions. 
For the superconducting order parameter correlations we obtain 
\bse
\label{eqs:2.15}
\bea
\langle{_r\Psi}_n^{\alpha}({\bf k})\,{_s\Psi}_m^{\beta}({\bf p})\rangle =
       \delta_{{\bf k},
   -{\bf p}}\,\delta_{n,m}\,\delta_{rs}\,\delta_{\alpha\beta}\,\frac{1}{2}\,
    {\cal M}_n({\bf k})\quad,
\nonumber\\
\label{eq:2.15a}
\eea
with
\be
{\cal M}_n({\bf k}) = \frac{1}{t  
+ \frac{\vert K^{(c)}\vert}{\delta{k_c}^2}\frac{1}{\ln{\frac{\Omega_0}{\vert\Omega_n\vert + 
{\bf k}^2/GH}}}}
\quad.
\label{eq:2.15b}
\ee
\ese%

Similarly, we find the fermionic propagators
\bse
\label{eqs:2.16}
\be
\langle{^i_rq}_{12}({\bf k})\,{^j_sq}_{34}({\bf p})\rangle =
   \delta_{{\bf k},-{\bf p}}\,\delta_{ij}\,\frac{G}{8}\,
   {^i_r\Gamma}_{12,34}^{(2)\,-1}({\bf k})\quad,
\label{eq:2.16a}
\ee
where
\bea
{^i_{0,3}\Gamma}_{12,34}^{(2)\,-1}({\bf k})&=&\delta_{13}\delta_{24}
   {\cal D}_{n_1-n_2}
   ({\bf k}) - \delta_{1-2,3-4}\delta_{\alpha_1\alpha_2}
      \delta_{\alpha_1\alpha_3}
\nonumber\\
&&\times 2\pi TGK^{(i)}\,{\cal D}_{n_1-n_2}({\bf k})\,
       {\cal D}^{(\rm i)}_{n_1-n_2}({\bf k})\ ,
\nonumber\\
\label{eq:2.16b}
\eea
and
\bea
{^0_{1,2}\Gamma}_{12,34}^{(2)\,-1}({\bf k})&=&-\delta_{13}\delta_{24}{\cal D}_{n_1-n_2}({\bf k})+\delta_{1+2,3+4}\delta_{\alpha_1\alpha_2}\delta_{\alpha_1\alpha_3}
\nonumber\\
&&\times 
\frac{4\pi TGK^{(c)}\,{\cal D}_{n_1-n_2}({\bf k})\,{\cal D}_{n_3-n_4}({\bf k})}{1 + 4 \pi TGK^{(c)} 
\ln{\frac{\Omega_0}{\vert\Omega_n\vert + 
{\bf k}^2/GH}}}
        \ .
\nonumber\\
\label{eq:2.16c}
\eea
Here ${\cal D}^{(\rm i)}$ is the spin-singlet propagator, which in
the limit of long wavelengths and small frequencies reads\cite{us_R}
\be
{\cal D}_{n}^{(\rm i)}({\bf k}) = \frac{1}{{\bf k}^2 + G(H+K^{(i)})\Omega_n}\quad.
\label{eq:2.16d}
\ee
\ese%

\subsection{Gaussian level renormalization group}
\label{subsec:III.B}

The standard momentum-shell renormalization group (RG) technique will be employed.\cite{Ma,Shankar,Fisher} We use $b$ as the RG length rescaling factor, and we rescale the wave number and two fields straightforwardly via
\bse
\label{eqs:3.1}
\bea
{\bm k}&\rightarrow&{\bm k'}/b\quad,
\label{eq:3.1a}\\
{\bm \Psi}_n({\bm k})&\rightarrow& b^{(2-\eta_{\Psi})/2} {\bm \Psi}^{\prime}_n({\bm k'})\quad,
\label{eq:3.1b}\\
q_{nm}({\bm k})&\rightarrow& b^{(2-\eta_q)/2} q_{nm}^{\prime}({\bm k'})\quad.  
\label{eq:3.1c}
\eea
The rescaling of imaginary time, frequency, or temperature is less
straightforward. In general, there are two
different time scales in the problem, namely, one that is associated
with the critical order-parameter fluctuations, and one that is associated
with the soft fermionic fluctuations. Therefore, we allow for two
different dynamical exponents, $z_{\Psi}$ and $z_q$. The temperature may then get rescaled via
\be
T \rightarrow b^{-z_M} T'\quad,
\label{eq:3.1d}
\ee
or via
\be
T \rightarrow b^{-z_q} T'\quad,
\label{eq:3.1e}
\ee
\ese%
How these various exponents should be chosen is discussed 
below.

In the tree, or zero-loop, approximation the RG equations for
the parameters in our field theory are determined as
\bse
\label{eqs:3.2}
\bea
t' &=& b^{2-\eta_{\Psi}}  t\quad,  
\label{eq:3.2a}\\
\frac{1}{G'H'T'_{\Psi}} &=& \frac{b^{-2}}{GHT_{\Psi}}\quad,  
\label{eq:3.2b}\\
\frac{1}{G'} &=& \frac{b^{-\eta_q}}{G}\quad,
\label{eq:3.2c}\\
H'T'_q &=& b^{2-\eta_q}  HT_q\quad,  
\label{eq:3.2d}\\
c'_1 T'^{1/2} &=& c_1 T^{1/2} b^{\frac{4-\eta_{\Psi}-\eta_q}{2}} \quad,  
\label{eq:3.2e}\\
c'_2 T'^{1/2} &=& c_2 T^{1/2} b^{\frac{-d+6-\eta_{\Psi}-2\eta_q}{2}} \quad,  
\label{eq:3.2f}
\eea
\ese%
Note that in giving Eqs.\ (\ref{eq:3.2e}) and (\ref{eq:3.2f}), the 
particular choice of $T$ was not yet specified because it is not obvious if a 
$z_q$ or a $z_{\Psi}$ should be used for these terms that describe a coupling 
between $q$ and $\Psi$ fields.

If we assume the Fermi-liquid degrees of freedom to be at a stable 
Fermi-liquid fixed point, we must choose $G$ and $H$ to be marginal, 
which implies 
\bse
\label{eqs:3.3}
\bea
\eta_q &=& 0\quad,  
\label{eq:3.3a}\\
z_{\Psi} &=& 2\quad,  
\label{eq:3.3b}\\
z_q &=& 2\quad.  
\label{eq:3.3c}
\eea
Here we find that two
dynamical exponents, $z_{\Psi}$ and $z_q$, have the same value, which is different from the 
ferromagnetic systems.\cite{us_paper_III}
We further choose
\be
\eta_{\Psi} = 2\quad,
\label{eq:3.3d}
\ee
which is implied within the logarithmic structure of Eq.\ (\ref{eq:2.15b}).
With these choices, we find that
\be
c'_2 = b^{\frac{-d+2}{2}} c_2\quad,
\label{eq:3.3e}
\ee
\ese%
As in the ferromagnetic case, there is a critical fixed
point where $c_1$ is marginal, and the fermions are diffusive, with exponents
given by Eqs.\ (\ref{eqs:3.3}). However, in contrast to the magnetic case, the
coupling constant $c_2$ of the term ${\cal A}_{\Psi-q^2}$ is RG irrelevant for all $d>2$, and
so are all higher terms in the expansion in powers of $q$. We therefore
conclude that the Gaussian critical behavior is exact.\cite{MSC} 
No additional logarithmic corrections exist here.
The most important 
technical difference that leads to the irrelevance of $c_2$ for this quantum phase transition, 
while for the quantum ferromagnetic transition it was marginal, is that the time scales
for the order-parameter fluctuations and the fermions, respectively, are the
same.\cite{add} This renders inoperative the mechanism that led the possibility of $c_2$
being marginal as in the ferromagnetic case. Physically, the very long range interaction
between the order-parameter fluctuations stabilizes the Gaussian
critical behavior. This is in agreement with the fact that long-ranged order
parameter correlations in classical systems stabilize mean-field critical
behavior.\cite{Fisher_Ma_Nickel_1972}

As noted above, Eq.\ (\ref{eq:3.3e}) implies that the Gaussian theory gave the exact critical
behavior. For completeness, the critical exponents, including logarithmic terms, are
\bse
\label{eqs:3.4}
\bea
\eta_{\Psi} &=& 2 - \frac{\ln{\ln{b^2}}}{\ln{b}}\quad,  
\label{eq:3.4a}\\
\nu &=& \frac{\ln{b}}{\ln{\ln{b^2}}}\quad,  
\label{eq:3.4b}\\
\gamma &=& 1\quad.  
\label{eq:3.4c}
\eea
\ese%
When $b\rightarrow\infty$ we have $\eta_{\Psi}=2$ and $\nu=\infty$.

\section{Conclusion}
\label{sec:IV}

We have investigated the quantum metal-superconductor phase transition in the present paper on the basis of an effective 
local field theory. With a simple renormalization group analysis, 
we have determined the critical behavior at 
the quantum metal-superconductor phase transition. 
In contrast to the disordered ferromagnetic case studied earlier, 
we showed that the previous results obtained with a nonlocal field theory were correct.
The reason is that the two
dynamical exponents, $z_{\Psi}$ and $z_q$, are exactly the same for
the disordered metal-superconductor quantum phase transition, which 
physically comes from the strong additional soft modes, or particle-hole excitations,
at zero temperature.

\acknowledgments

This work was 
supported by the NSF under grant numbers DMR-99-75259 and DMR-01-32726.

\end{document}